\documentstyle[12pt,aasms4]{article}
\def\be{\begin{equation}}
\def\ee{\end{equation}}
\def\ba{\begin{eqnarray}}
\def\ea{\end{eqnarray}}
\def\etal{et al.}

\def\la{\mathrel{\mathpalette\fun <}}

\def\fun#1#2{\lower3.6pt\vbox{\baselineskip0pt\lineskip.9pt
        \ialign{$\mathsurround=0pt#1\hfill##\hfil$\crcr#2\crcr\sim\crcr}}}

\slugcomment{To be submitted to ApJ}

\begin{document}
\null\vspace{-62pt}
\begin{flushright}
\today
\end{flushright}

\title{Model Independent Primordial Power Spectrum \\
from Maxima, Boomerang, and DASI Data}
\author{Yun Wang$^1$, \& Grant J. Mathews$^2$}
\affil{{$^1$\it Department of Physics \& Astronomy} \\
{\it University of Oklahoma, Norman, OK 73019\\}
{\it email: wang@mail.nhn.ou.edu}}
\affil{{$^2$\it Center for Astrophysics, Department of Physics} \\
{\it University of Notre Dame, Notre Dame, IN 46556\\}
{\it email: gmathews@nd.edu}}

\vspace{.4in}
\centerline{\bf Abstract}
\begin{quotation}

A model-independent determination of the primordial power spectrum of matter density 
fluctuations could uniquely probe physics of the very early universe, and
provide powerful constraints on inflationary models.
We parametrize the primordial power spectrum $A_s^2(k)$ as an arbitrary function, 
and deduce its binned amplitude from the cosmic microwave background radiation
anisotropy (CMB) measurements of Maxima, Boomerang, and DASI.
We find that for a flat universe with $A_s^2(k)=1$ (scale-invariant)
for scales $k<0.001\,$h/Mpc, the primordial power spectrum 
is marginally consistent with a scale-invariant Harrison-Zeldovich spectrum.
However, we deduce a rise in power compared to a scale-invariant power spectrum
for $0.001\,h/\mbox{Mpc} \la k \la 0.01\,h/\mbox{Mpc}$.
Our results are consistent with large-scale structure data, and seem to
suggest that the current observational data allow for the possibility of
unusual physics in the very early universe.

\end{quotation}


\section{Introduction}

The inflationary paradigm is presently the most plausible solution to 
the problems of the standard cosmology.  Inflation is consistent 
with all current observational data. However, we are still far 
from establishing a definitive model of inflation.
There exists a broad range of inflationary models
(cf.~\cite{Kolb96,Turner97}), many of which appear consistent with
observational data.
In order to quantify what can be known about inflation one desires
model-independent measurements of 
the primordial power spectrum of matter density fluctuations as these
can provide unique and powerful constraints on inflationary models.

Although it has been conventional to take the primordial power
spectrum to be a featureless power law in analyzing cosmological
data, there are both theoretical and observational reasons to allow
the primordial power spectrum to be a free function.
That is, some inflationary models predict power spectra that are
almost exactly scale-invariant (\cite{Linde83}),
or are described by a power law with spectral index less than one
(\cite{naturalinf,extendedinf}), 
while others predict power spectra
with slowly varying spectral indices (\cite{Wang94}), or
with broken scale invariance (\cite{Holman91ab,Randall96,Adams97,Les97,Les99}).
The latter represents unusual physics in the very early universe.
For example, inflation might occur in multiple stages
in effective theories with two scalar fields (\cite{Holman91ab}), 
or in a succession of short bursts due to
symmetry breaking during an era of inflation in 
supergravity models (\cite{Adams97}).

There is also tentative observational evidence for a peak in the power
spectrum of galaxies at $k\sim 0.05\,$Mpc$^{-1}$ 
(\cite{Einasto97,Baugh98,Retslaff98,Broad99,Gramann99,Gramann00}).
The simplest explanation for such a peak in the galaxy power spectrum
is a new feature in the primordial power spectrum.

The cosmic microwave background radiation anisotropies (CMB) are signatures of the 
primordial matter density fluctuations and gravity waves 
imprinted at the time when photons decoupled 
from matter. The large-scale structure in the distribution of galaxies is a 
direct consequence of the power spectrum of the primordial density fluctuations. 
Wang, Spergel, \& Strauss (1999) have explored how one can use the upcoming CMB 
data from the Microwave Anisotropy Probe 
(MAP; \cite{MAP}; {\tt http://map.gsfc.nasa.gov})
and the large-scale structure 
data from the the Sloan Digital Sky Survey 
(SDSS; cf., \cite{SDSS})
to obtain a model-independent measurement of the
primordial power spectrum, and to extract simultaneously
the cosmological parameters.

In this paper, we implement the concept of a model-independent measurement
of the primordial power spectrum from Wang et al. (1999) to extract
cosmological information from the CMB data from Maxima (\cite{Maxima}),
Boomerang (\cite{Boomerang}), and DASI (\cite{Halverson01}).
We parametrize the primordial power spectrum as a continuous
and arbitrary function determined by its amplitude at several wavenumbers
[which are equally spaced in $\log(k)$] via linear interpolation.
We then measure these ``binned'' amplitudes from
the CMB data of Maxima, Boomerang, and DASI.

\section{Measurement of the Primordial Power Spectrum}

We parametrize the primordial power spectrum as
\ba
\label{eq:As2(k)}
A_s^2(k)&=& \left(\frac{k_i -k}{k_{i}-k_{i-1}}\right)\,a_{i-1}+
		\left(\frac{k-k_{i-1}}{k_{i}-k_{i-1}}\right)\,a_{i},
		\hskip 1cm k_{i-1} < k \le k_i, \,\, i=1,n	\nonumber \\
A_s^2(k)&=& a_0=1, \hskip 1cm k\le k_0=k_{min},  		\nonumber \\
A_s^2(k)&=& a_n, \hskip 1cm k\ge k_n =k_{max},	
\ea
where 
\[
k_i=\exp\left[ \frac{i}{n}\ln\left(\frac{k_n}{k_0}\right)+ \ln(k_0)\right],
\hskip 1cm i=0,n
\]
We impose $A_s^2(k)=1$ for $k\leq k_0=k_{min}$ for two reasons.
First, on the largest scales, the CMB data is consistent with a scale-invariant
primordial power spectrum (\cite{COBE1,COBE2}), i.e., $A_s^2(k)=1$. 
Second, the bin amplitude of the
primordial power spectrum on the largest scales is poorly constrained by the CMB 
data due to the effects of cosmic variance. We also assume that $A_s^2(k)=a_n$ for 
$k> k_n = k_{max}$, because
$k_{max}$ is close to the scale corresponding to the angular resolution
of Maxima and Boomerang. We choose $k_{min}=0.001\,h$/Mpc and
$k_{max}=0.1\,h$/Mpc.

We perform the parameter estimation by computing the CMB angular power spectrum
$C_l({\bf s})$ for a discrete set of cosmological parameters denoted as {\bf s}. 
To save computational time and storage space, we have assumed a flat universe
and limited our parameter search to a grid
in six cosmological parameters, $\{H_0, \Omega_m, \Omega_b, a_1, a_2, a_3\}$,
with $\Omega_{\Lambda}=1-\Omega_m$,  $\tau_{ri}=0$. The 
$a_i$ ($i=1,3 $) parametrize the primordial power spectrum $A_s^2(k)$
as in Eq.(\ref{eq:As2(k)}) with $n=3$.
Our assumption of a flat universe is consistent with all current observational
data, and preferred by non-fine-tuned inflationary models.
We use the COBE normalization for all the theoretical models.

Following the Maxima, Boomerang, and DASI teams 
(\cite{Jaffe00,Lange00,Pryke01}), we
use an offset lognormal likelihood function to define a $\chi^2$ 
goodness of fit (\cite{Bond00}). That is, for a given theoretical model
with ${\cal C}_l({\bf s})\equiv l(l+1) C_l({\bf s})/(2\pi)$, we define
\ba
\label{eq:chi2}
&&\chi^2({\bf s}) \equiv -2\, \ln {\cal L}
= \chi^2_d + \chi^2_{cal}+ 
\chi^2_{beam}, \\
\label{eq:chi2d}
&&\chi^2_d=\sum_{i,j} (Z_i^t - Z_i^d) M_{ij}^Z (Z_j^t - Z_j^d),\\
&&\chi^2_{cal}= \sum_{\alpha} \frac{ (u_{\alpha}-1)^2}{ \sigma_{\alpha}^2},
\ea
where
\ba
&&Z_i^d  \equiv \ln (D_i +x_i)  \\
&&Z_i^t  \equiv \ln \left(u_{\alpha} \sum_{l_{min}^{i}}^{l_{max}^{i}} f_{il} 
{\cal C}_l \, g_l +x_i \right), \\
&&f_{il}= \frac{ W^i_l/l}{\sum_{l_{min}^{i}}^{l_{max}^{i}} W^i_l/l},
\ea
and
\be
M_{ij}^Z= M_{ij} (D_i+x_i)(D_j+x_j).
\label{eq:matr}
\ee
In the above equation, $D_i$ is the measured CMB bandpower in the $i$th bin, 
$x_i$ is an offset which depends upon experimental details, $W_l = B_l^2$ is the
experimental window function (where $B_l$ is the beam function).
The weight matrix $M_{ij}$ is given by the Fisher matrix
$F_{ij} = - \frac{\partial^2 {\cal L}}{\partial {\cal C}_i
\partial {\cal C}_j }$. The calibration uncertainty is parametrized by $u_{\alpha}$,
the factor of overall relative calibration;
and which has a dispersion of $\sigma_{\alpha}=0.08$ and 
0.2 for Maxima-1 and Boomerang, respectively.
We also consider beam uncertainties for Boomerang.
Following Lange et al. (2000), we parametrize the beam uncertainty 
with $\exp\left\{ -(l+0.5)^2 [\Delta(\theta_s^2)] \right\}$,
with $\langle \Delta(\theta_s^2) \rangle = 2 \theta_{FWHM} \,
\Delta \theta_{FWHM}=(572.0)^{-2}$, where we have used
$\theta_{FWHM}=12.9'$, and $\Delta \theta_{FWHM}=1.4'$. Hence, we write
\be
\chi^2_{beam}= \sum_i \frac{ \left( \sum_{l_{min,i}}^{l_{max,i}}
f_{il} {\cal C}_l \exp\left[ - g(l) \, \delta_{beam}\right] - 
\sum_{l_{min,i}}^{l_{max,i}}
f_{il} {\cal C}_l \right)^2 }
{\left( \sum_{l_{min,i}}^{l_{max,i}}
f_{il} {\cal C}_l \exp\left[- g(l) \right] 
- \sum_{l_{min,i}}^{l_{max,i}}
f_{il} {\cal C}_l \right)^2 },
\ee
where we have defined $\delta_{beam} \equiv 572.0^2\,\Delta(\theta_s^2)$,
and $g(l)\equiv [(l+0.5)/572.0]^2$.

We use the higher precision Maxima and Boomerang data as published
in {\cite{Lee01}} and {\cite{Netterfield01}}. The relevant 
experimental details (beam functions, covariance matrix, 
and log-normal offsets $x_i$) have not yet been released.
It has been shown that in the absence of the knowledge of
$x_i$, Gaussian statistics describe the data better than 
a log-normal distribution with arbitrary $x_i$ (\cite{Bond00}).
Therefore, we replace Eq.(\ref{eq:chi2d}) with
\be
\label{eq:chi2G}
\chi^2({\bf s}) =\sum_i \frac{ \left[{\cal C}_{data}^i - {\cal C}_{BP}^i ({\bf s})\right]^2}
{\sigma_i^2},
\ee
where ${\cal C}_{data}^i$ are the experimental band-powers with measurement errors $\sigma_i$, 
and
\be
{\cal C}_{BP}^i({\bf s})= \frac{ 
\sum_{l_{min,i}} ^{l_{max,i} } 
W_l {\cal C}_l ({\bf s})/l }{\sum_{l_{min,i}} ^{l_{max,i}} W_l /l },
\hskip 2cm
W_l=B_l^2= e^{-(0.425 \theta_{FWHM} l)^2},
\ee
where $\theta_{FWHM}=10'$ and $12.9'$ for Maxima and Boomerang
respectively.
Eq.(\ref{eq:chi2G}) assumes symmetric error bars on ${\cal C}_{data}^i$. For the asymmetric 
errors given by Maxima, we take $\sigma_i=\sigma_{i,-}$ for 
${\cal C}_{data}^i > {\cal C}_{BP}^i ({\bf s})$, and $\sigma_i=\sigma_{i,+}$ otherwise.

The DASI team has released the experimental details (window functions
$f_{il}$, log-normal offsets $x_i$, and covariance matrix $V_{ij}$)
together with their data ({\tt http://astro.uchicago.edu/dasi/}). 
This enables us to use log-normal statistics 
[Eqs.(\ref{eq:chi2d})-(\ref{eq:matr})], with the weight matrix
$M_{ij}$ given by the inverse of the matrix
$N_{ij}=V_{ij}+ \sigma_{\alpha} D_i D_j$, where 
$\sigma_{\alpha}=0.08$ is the calibration uncertainty of DASI.

For a given set of cosmological parameters, 
$\{H_0, \Omega_m, \Omega_b, a_1, a_2, a_3\}$, we marginalize
the model over the calibration uncertainties of Maxima,
Boomerang, and DASI, and over the beam uncertainty of Boomerang.
The beam uncertainty is only marginalized for Boomerang, since
only this experiment appears to have significant beam uncertainty.

\section{Results}

Since only five values of $H_0= 100\,h\,$km/s$\,$Mpc$^{-1}$
($h=0.5$, 0.6, 0.7, 0.8, 0.9) are computed for our grid, we present our results
for the different values of $h$ separately, instead of marginalizing
over $h$.

Figs.1(a)-(c) show the data from (a) Maxima, (b) Boomerang, and
(c) DASI, together with best-fit models.
In each figure, the solid curve is the best-fit model to the
combined Maxima, Boomerang, and DASI data, while the dotted line
is the best-fit model to the (a) Maxima, (b) Boomerang, and
(c) DASI data separately. The dotted error bars are the errors 
on each data point including calibration uncertainty.
We have combined the data from 
Maxima, Boomerang, and DASI assuming that the three experiments 
are independent of each other.

Table 1 shows the best-fit models to the combined Maxima, 
Boomerang, and DASI data.
\begin{table}[htb]
\caption{Best fit models to the combined Maxima, 
Boomerang, and DASI data}
\begin{center}
\begin{tabular}{ccccccccccc}
\hline\hline
      $h$  & $\Omega_m$ &$\Omega_b$& $a_1$ &$a_2$&$a_3$& 
	$u_{mx}$ &  $u_{Boom}$ & $u_{DASI}$ & $\delta_{beam}$
	& $\chi^2_{min}$\\ 
\hline
0.5 &  0.700 & 0.070 & 2.5 & 1.7 & 1.3 & 0.933 & 1.008 & 0.952 & 0.10 & 28.4500 \\
0.6 &  0.425 & 0.055 & 2.0 & 1.2 & 1.0 & 0.974 & 1.056 & 0.987 & 0.05 & 28.2064 \\
0.7 &  0.250  &  0.040 & 1.5 & 0.9 & 0.9 & 0.978 & 1.056 &  0.994 & 0.05  & 29.3004\\
0.8 &  0.200 &  0.040 & 1.0 & 0.8 & 1.0 & 0.958 & 1.048 &  0.971 & 0.10 & 29.8050\\
0.9 &  0.120 & 0.030 & 0.9 & 0.7 & 1.0 & 0.981 & 1.048 &  0.990 &  0.05 & 29.0687  \\
\hline
\end{tabular}
\end{center}
\end{table}

Table 2-4 show the best-fit models to the Maxima, 
Boomerang, and DASI data separately.
\begin{table}[htb]
\caption{Best fit models to the Maxima data}
\begin{center}
\begin{tabular}{cccccccc}
\hline\hline
      $h$  & $\Omega_m$ &$\Omega_b$& $a_1$ &$a_2$&$a_3$& 
	$u_{mx}$ & $\chi^2_{min}$\\ 
\hline
0.5 &   0.800 &    0.100 & 3.0 & 1.5 & 1.5 & 0.997  &  4.3313\\
0.6 &   0.600 &   0.100 & 2.5 & 1.0 & 1.5 & 0.997 &  3.2590\\
0.7 &   0.400 &  0.080 & 0.9 & 0.6 & 1.0 &0 1.010 & 3.9188\\
0.8 &   0.300 &  0.070 & 1.0 & 0.6 & 1.2 & 0.994 & 4.5866 \\
0.9 &   0.200 &    0.060 & 0.7 & 0.5 & 1.3 & 1.000 & 4.7581 \\
\hline
\end{tabular}
\end{center}
\end{table}

\begin{table}[htb]
\caption{Best fit models to the Boomerang data}
\begin{center}
\begin{tabular}{ccccccccc}
\hline\hline
      $h$  & $\Omega_m$ &$\Omega_b$& $a_1$ &$a_2$&$a_3$& 
	$u_{B00}$ & $\delta_{beam}$ & $\chi^2_{min}$\\ 
\hline
0.5 &  0.600 &  0.050 & 5.0 & 3.0 & 1.5 & 0.968 & -0.05 & 11.4684\\
0.6 &  0.375 &   0.050 & 2.5 & 1.5 & 1.0 & 0.968 &  0.00 & 11.7898 \\
0.7 &  0.250 &  0.040 & 2.0 & 1.2 & 0.9 & 0.984 &  0.00 & 12.1868 \\
0.8 &  0.150 &  0.030 & 1.5 & 0.9 & 0.8 & 1.016 &  0.00  & 12.4757 \\
0.9 &  0.120 &  0.030 & 0.9 & 0.8 & 0.9 & 1.008 &  0.05 & 13.5643 \\
\hline
\end{tabular}
\end{center}
\end{table}

\begin{table}[htb]
\caption{Best fit models to the DASI data}
\begin{center}
\begin{tabular}{cccccccc}
\hline\hline
      $h$  & $\Omega_m$ &$\Omega_b$& $a_1$ &$a_2$&$a_3$& 
	$u_{DASI}$ & $\chi^2_{min}$\\ 
\hline
0.5 & 0.800 &  0.070 &  2.5 & 1.7 & 1.3 & 0.990 & 4.3187\\
0.6 & 0.450 &  0.050 & 4.0 & 1.7 & 1.5 & 1.010 & 3.3240 \\
0.7 & 0.275 & 0.040 & 3.0 & 1.2 &1.3 & 1.006 & 2.5608\\
0.8 & 0.200 &   0.040 & 2.0 & 0.9 & 1.3 & 0.994 & 2.6985 \\
0.9 & 0.120 &  0.030 & 0.9 & 0.6 & 1.0 & 1.029 & 2.3055\\
\hline
\end{tabular}
\end{center}
\end{table}

We constrain parameters individually by marginalizing over all other parameters.
Figs.2(a)-(e) show the likelihood functions for the combined Maxima, Boomerang,
and DASI data for the set of parameters $\{\Omega_m, \Omega_b, a_1, a_2, a_3\}$,
for $h=0.5$ (dot), 0.6 (solid), 0.7 (dashed), 0.8 (long-dashed),
and 0.9 (dot-dashed).
The likelihood functions have been derived using $\chi^2$ values which resulted
from multi-dimensional interpolations of the $\chi^2$ values computed for the
grid of models (cf.~\cite{Tegmark00a}).

Fig.3 shows the primordial power spectrum $A_s^2(k)$ measured from 
the combined Maxima, Boomerang, and DASI data 
for $h=0.6$ (solid), and 0.7 (dotted).
The $\pm$1$\sigma$ errors are deduced from Fig.2(c)-(e).

Note that $A_s^2(k)=1$ corresponds to the scale-invariant Harrison-Zeldovich 
$n_s=1$ spectrum, with the primordial scalar power spectrum conventionally 
defined as $k\,A_s^2(k) \propto k^{n_s}$. Clearly, the primordial power spectrum 
is marginally consistent with a scale-invariant Harrison-Zeldovich spectrum.
However, there is a rise in the power at 
$0.001\,h\,{\mbox{Mpc}}^{-1} \la k \la 0.01\,h\, {\mbox{Mpc}}^{-1}$ 
compared to a scale-invariant power spectrum.
Although this rise is not statistically convincing, it could be
an indication of interesting physics worthy of future investigation.

\section{Discussion}

Due to parameter degeneracies between $h$ and $\Omega_m$ and $\Omega_b$, 
the CMB data constrain $h$ only weakly. Fortunately, there are a number
of independent methods for constraining $h$ (\cite{RobK,DavidB,lensH0}).
The current combined CMB data from Maxima, Boomerang, and DASI marginally
prefer $h=0.6$ (see Table 1).

Although the CMB data is sensitive to $\Omega_m h^2$ and $\Omega_b h^2$, the
likelihood curves for different values of $h$ do not overlap [see 
Figs.2(a)(b)], since these curves correspond to different values
of $\Omega_{\Lambda}$ (we assume that $\Omega_{\Lambda}=1-\Omega_m$).
The likelihood curves for $\Omega_b h^2$ are less sensitive to
the value of $h$ [see Fig.2(b)].
This is because varying $\Omega_b$ has a much smaller effect
on $\Omega_{\Lambda}$ than varying $\Omega_m$.

Our measurement of the primordial power spectrum [see Fig.3]
is consistent with the large-scale structure data which seem to
indicate a peak in the matter power spectrum at $k\sim 0.05\,$Mpc$^{-1}$ 
(\cite{Einasto97,Baugh98,Retslaff98,Broad99,Gramann99,Gramann00}).
We note that the real space power spectrum of the PSCz Survey, 
from 0.01 to 300 h/Mpc (\cite{Hamilton00}), does not show this peak.
The data from 2df (\cite{2df}) and SDSS (\cite{SDSS})
should help clarify any features in the large-scale structure power 
spectrum. However, if there is a feature at $k\la 0.002\,h\,$Mpc$^{-1}$
(corresponding to the characteristic length scale of the SDSS, see 
\cite{Wang99}), satelite CMB data from MAP or Planck will be required
to constrain such features in the primordial power spectrum (\cite{Wang99}).

A number of authors have used Maxima and Boomerang data to derive cosmological 
constraints 
(\cite{Abazajian00,Amendola00,Avelino00,Balbi00,Bento00,Bouchet00,Brax00,Bridle00,Contaldi00,Durrer00,Enqvist00,Esposito00,Griffiths00,Hannestad00,Hannestad00b,Hu00,Jaffe00,Kanazawa00,Kinney00,Landau00,Lange00,Lesgourgues00,McGaugh00,Melchiorri00,Pad00,Tegmark00b,Tegmark00c,White00}).
Our work is unique in allowing the primordial power spectrum to be an arbitrary 
function, thus allowing the possibility for detecting new
features in the primordial power spectrum.

Our results seem to indicate that the current observational data do not rule out 
unusual physics (such as multiple-stage inflation) in the very early universe.
The upcoming data from the CMB satellite missions MAP ({\cite{MAP}) and 
Planck (\cite{Planck}), and the large-scale structure data
from 2df (\cite{2df}) and SDSS (\cite{SDSS}) should allow for a 
more definitive measurement of the primordial power spectrum (\cite{Wang99}).
These data will more precisely constrain the possibility for such complex 
physics in the very early universe.

\vskip .2in
\centerline{\bf Acknowledgements}
We acknowledge the use of CMBFAST (\cite{SeljakZ96}) in computing the theoretical models.
It is a pleasure for us to thank the referee for helpful suggestions.
Work supported in part by NSF CAREER grant AST-0094335 at the Univ. of Oklahoma,
and DOE grant DE-FG02-95ER40934 at Univ. of Notre Dame.

\clearpage

\setcounter{figure}{0}
\figcaption[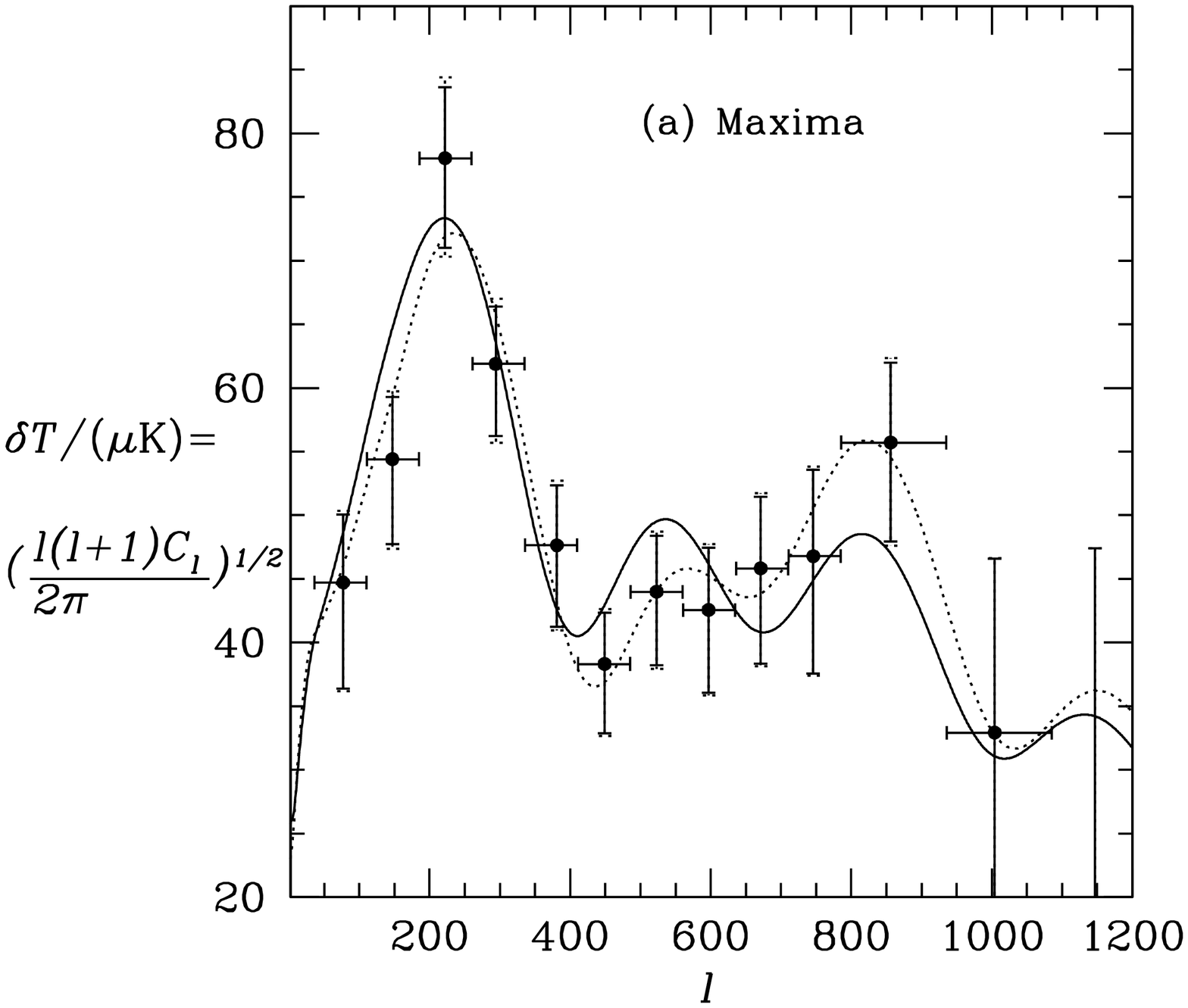]{The data from (a) Maxima, (b) Boomerang, and
(c) DASI, together with best-fit models.
In each figure, the solid curve is the best-fit model to the
combined Maxima, Boomerang, and DASI data, while the dotted line
is the best-fit model to the (a) Maxima, (b) Boomerang, and
(c) DASI data separately.
The dotted error bars are the errors 
on each data point including calibration uncertainty.}

\figcaption[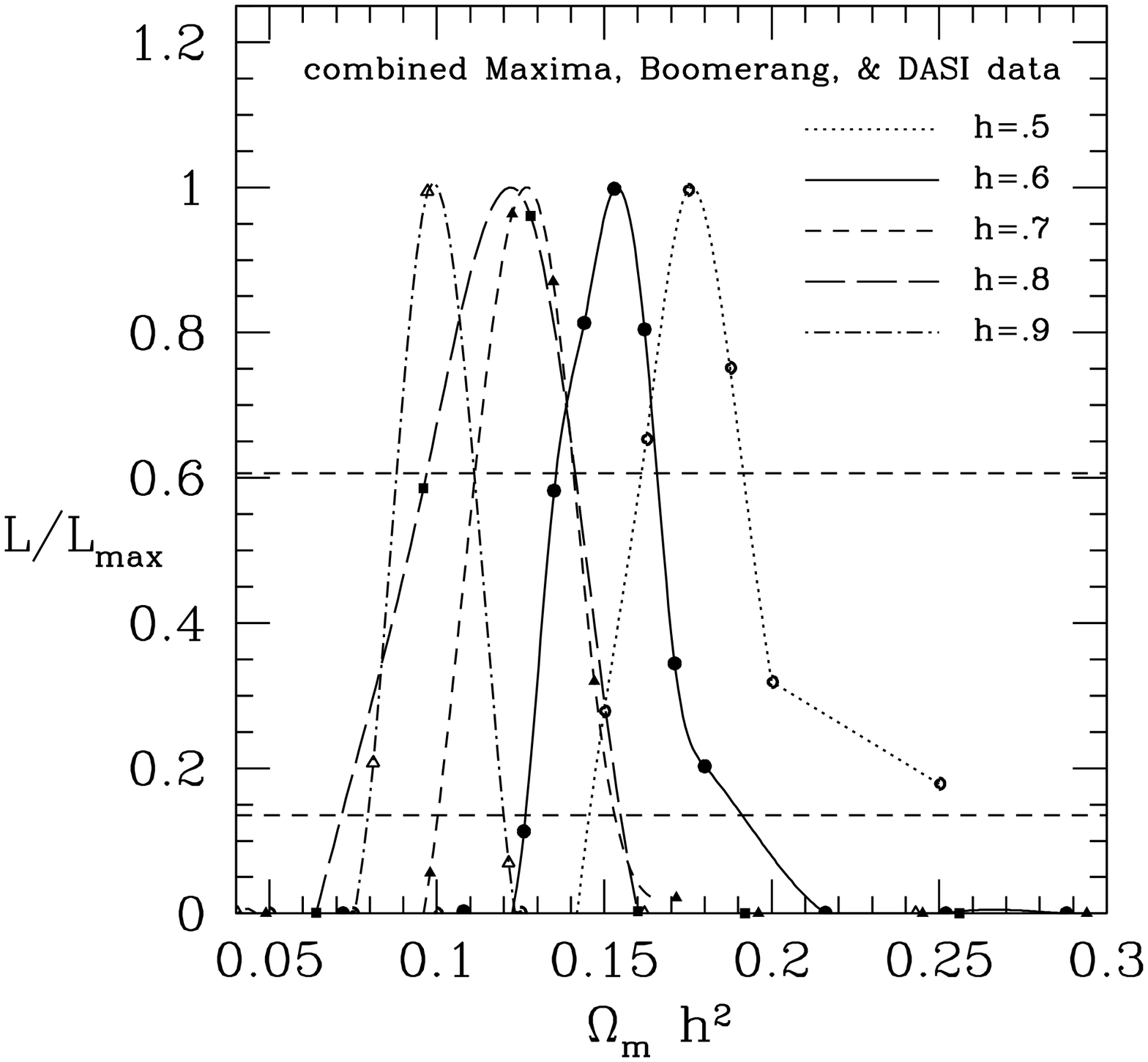]{The likelihood functions for the combined Maxima, Boomerang,
and DASI data for the set of parameters $\{\Omega_m, \Omega_b, a_1, a_2, a_3\}$,
for $h=0.5$ (dot), 0.6 (solid), 0.7 (dashed), and 0.8 (long-dashed).}

\figcaption[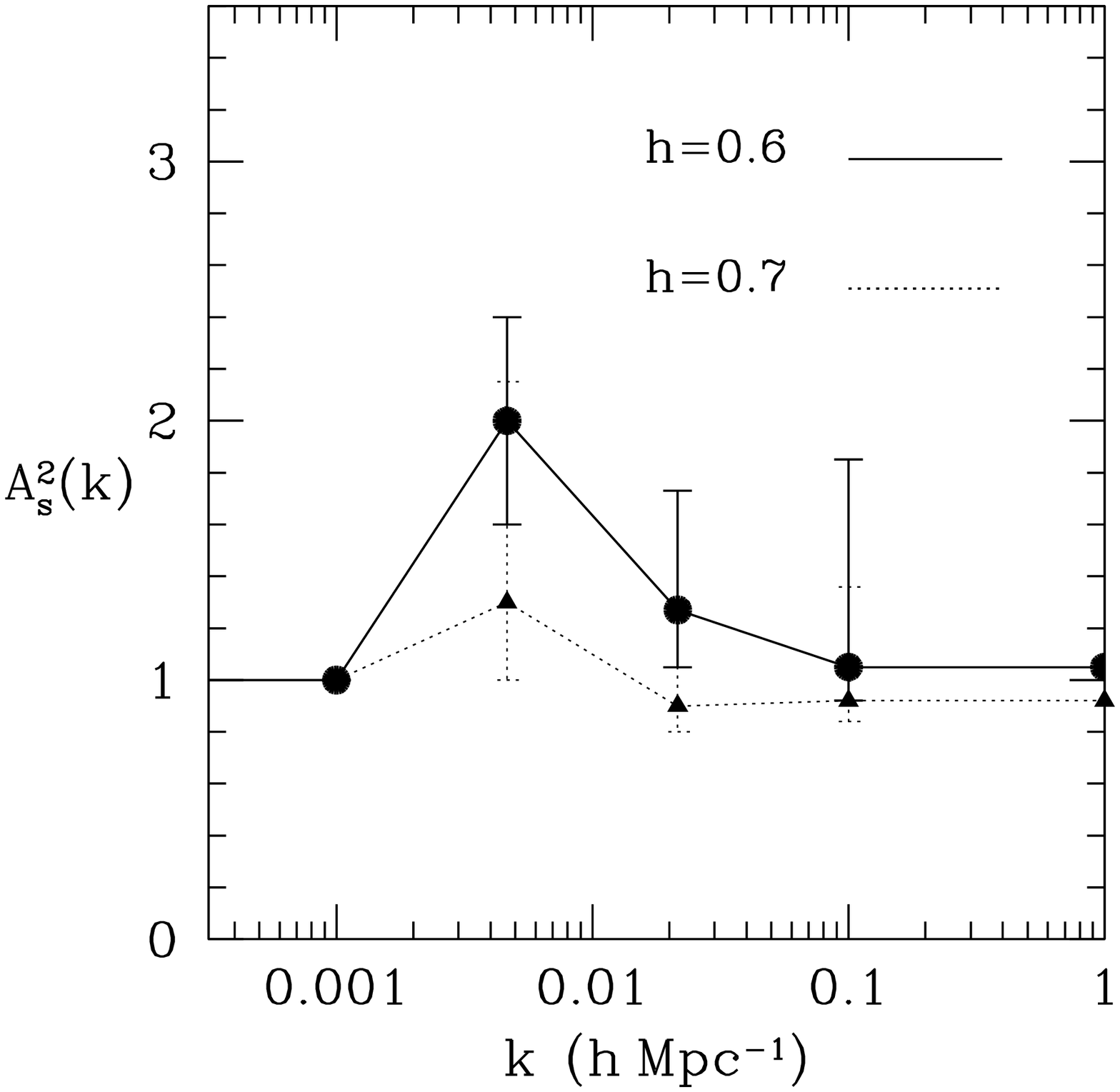]{The primordial power spectrum $A_s^2(k)$ measured from 
the combined Maxima, Boomerang, and DASI data 
for $h=0.6$ (solid), and 0.7 (dotted).
The $\pm$1$\sigma$ errors are estimated from Fig.2(c)-(e).}

\clearpage

\setcounter{figure}{0}
\plotone{f1a.eps}
\figcaption[f1a.eps]{The data from (a) Maxima, (b) Boomerang, and
(c) DASI, together with best-fit models.
In each figure, the solid curve is the best-fit model to the
combined Maxima, Boomerang, and DASI data, while the dotted line
is the best-fit model to the (a) Maxima, (b) Boomerang, and
(c) DASI data separately.
The dotted error bars are the errors 
on each data point including calibration uncertainty.}

\plotone{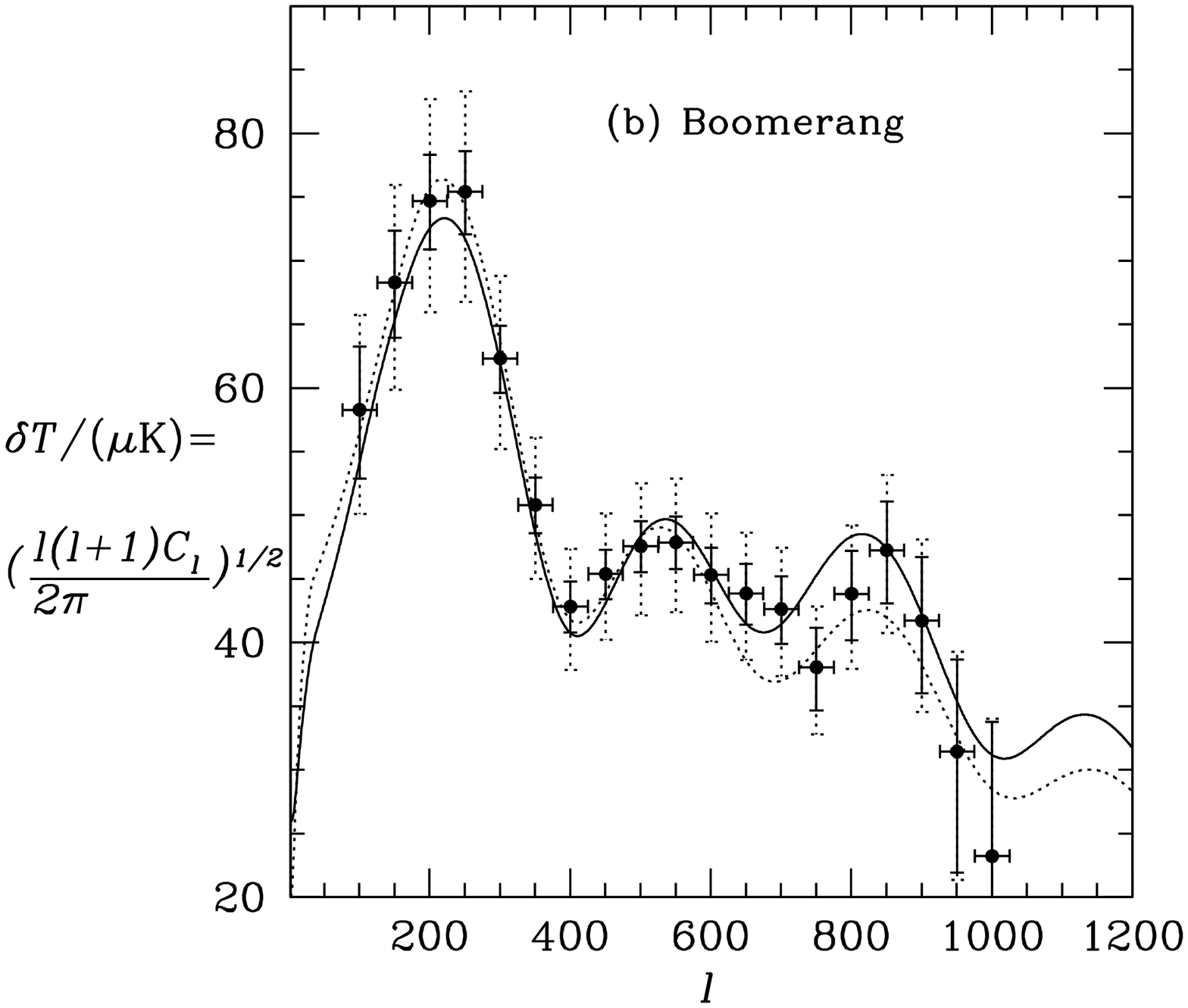}
\setcounter{figure}{0}
\figcaption[f1b.eps]{ (b) Boomerang data.}

\plotone{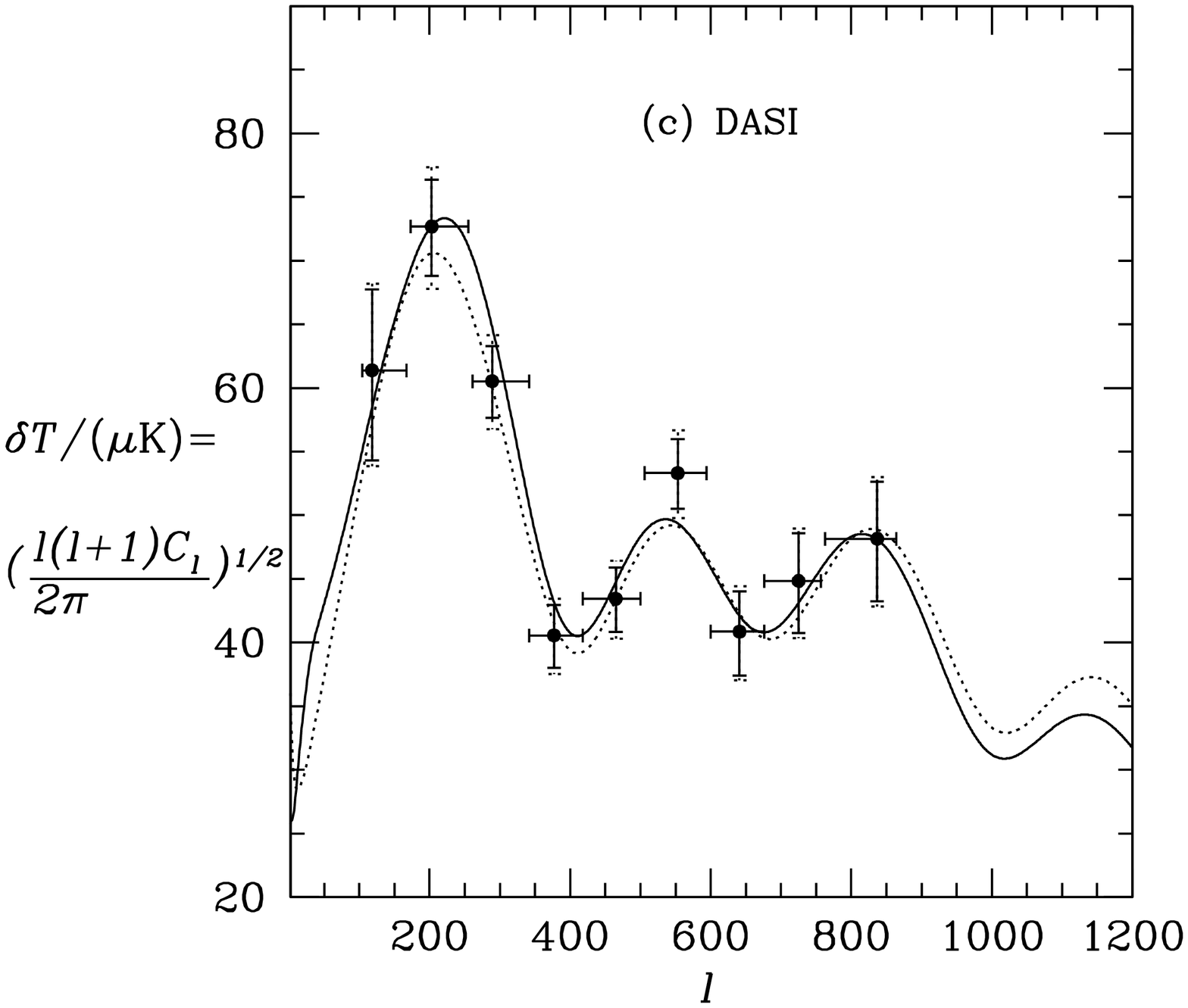}
\setcounter{figure}{0}
\figcaption[f1c.eps]{ (c) DASI data.}

\plotone{f2a.eps}
\figcaption[f2a.eps]{The likelihood functions for the combined Maxima, Boomerang,
and DASI data for the set of parameters $\{\Omega_m, \Omega_b, a_1, a_2, a_3\}$,
for $h=0.5$ (dot), 0.6 (solid), 0.7 (dashed), and 0.8 (long-dashed).
(a) $\Omega_m h^2$. }

\setcounter{figure}{1}
\plotone{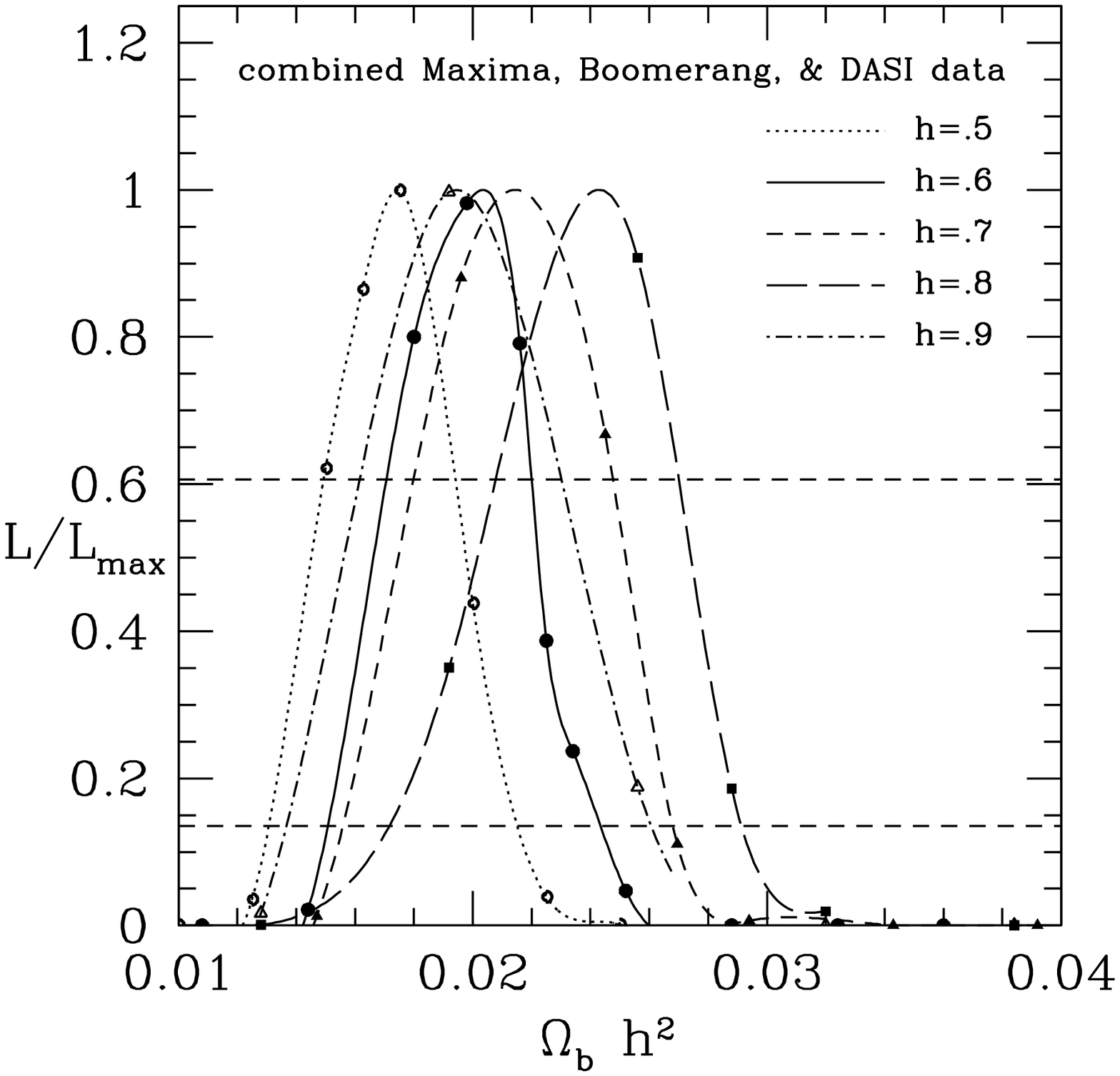}
\figcaption[f2b.eps]{
(b) $\Omega_b h^2$.}

\setcounter{figure}{1}
\plotone{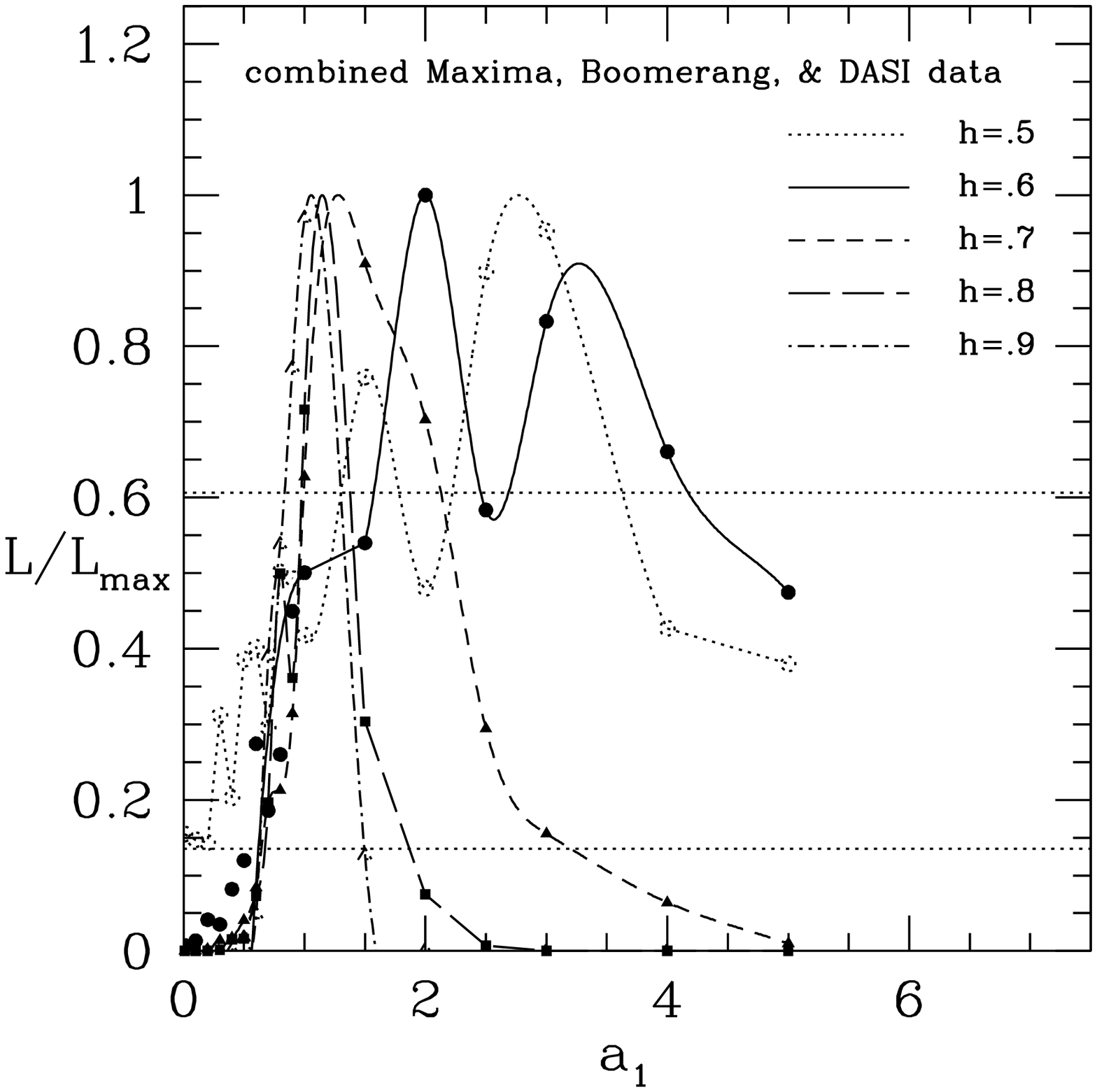}
\figcaption[f2c.eps]{
(c) $a_1$.}

\setcounter{figure}{1}
\plotone{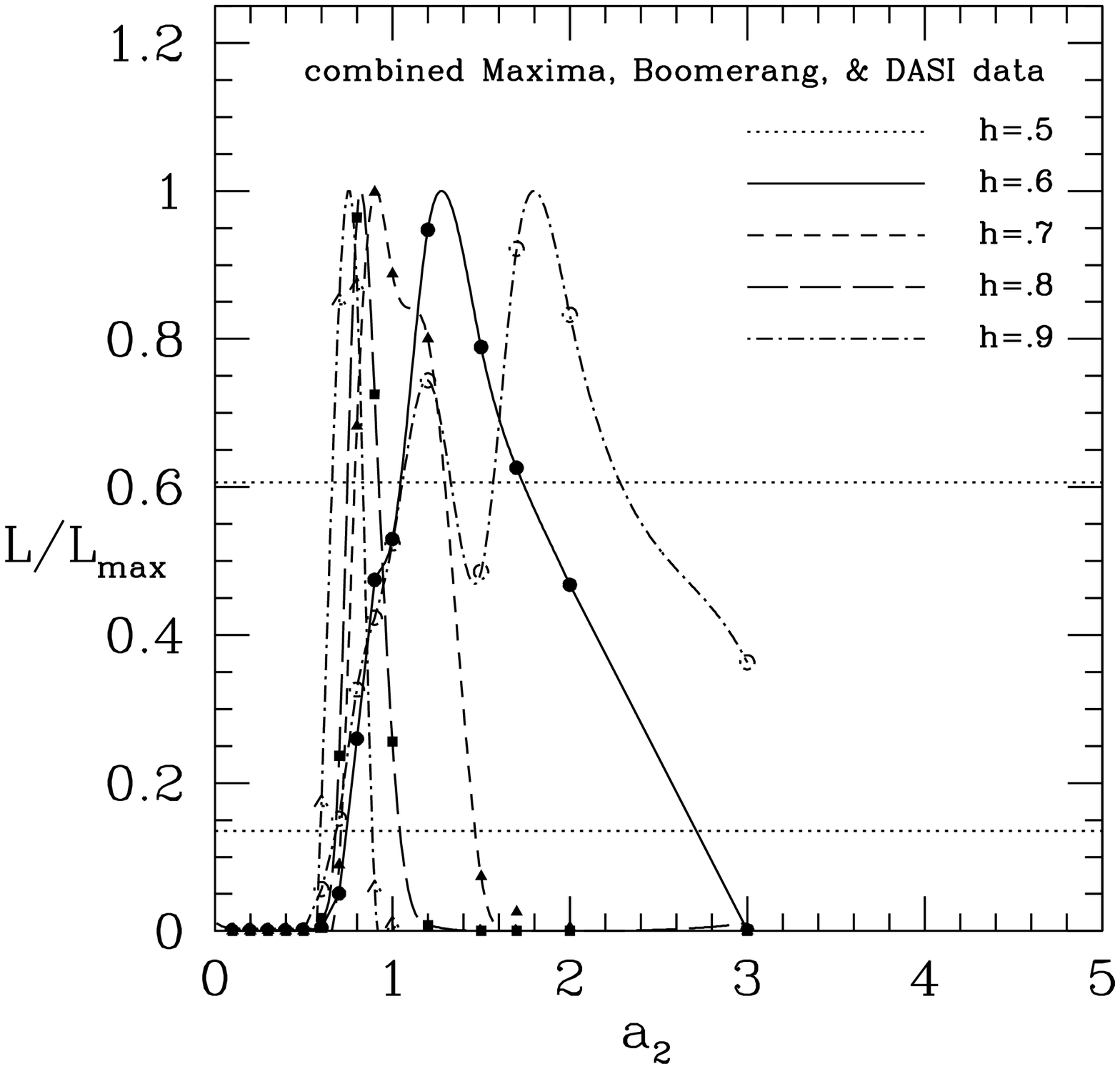}
\figcaption[f2d.eps]{
(d) $a_2$.}

\setcounter{figure}{1}
\plotone{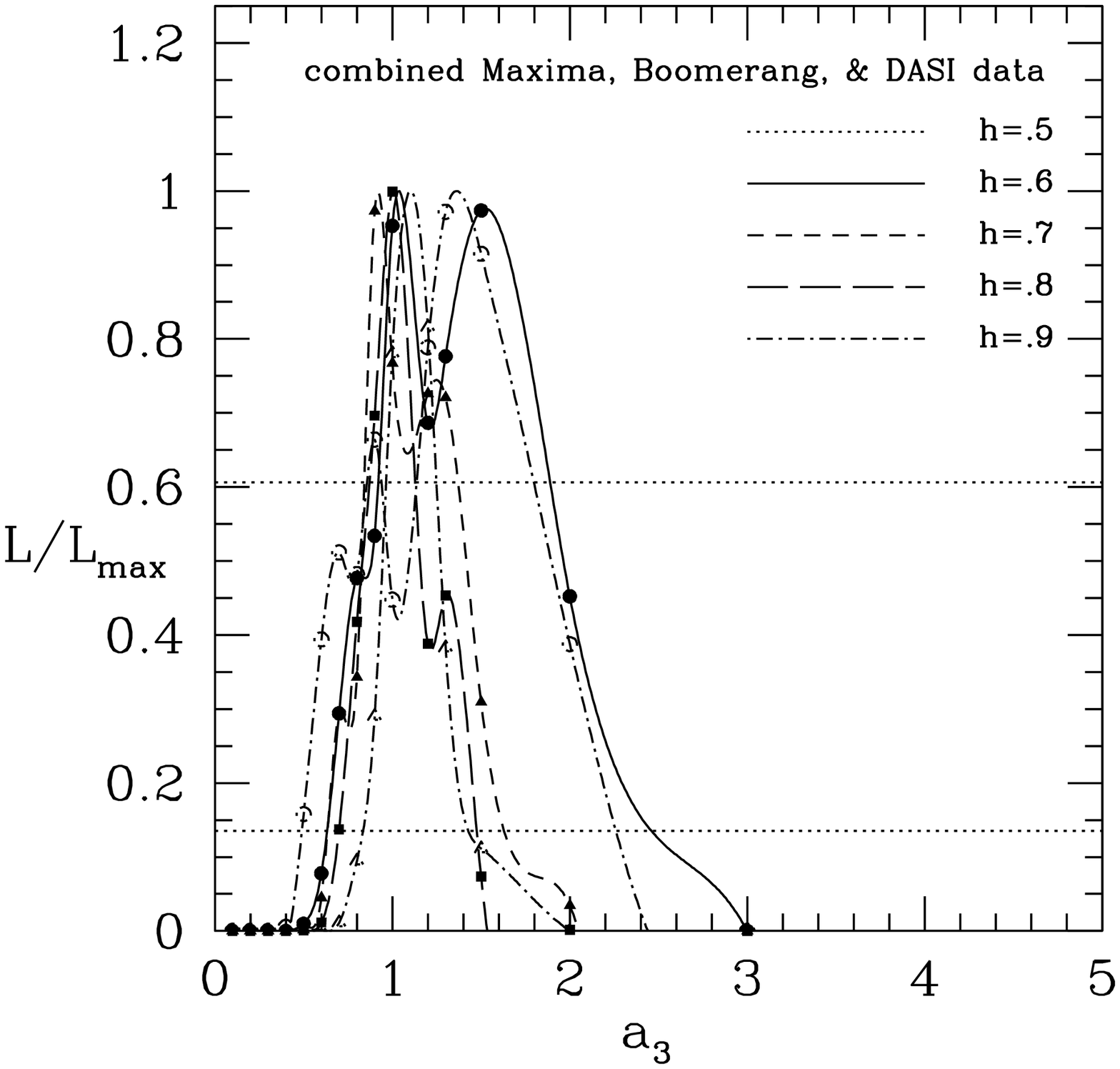}
\figcaption[f2e.eps]{
(e) $a_3$.}

\plotone{f3.eps}
\figcaption[f3.eps]{The primordial power spectrum $A_s^2(k)$ measured from 
the combined Maxima, Boomerang, and DASI data 
for $h=0.6$ (solid), and 0.7 (dotted).
The $\pm$1$\sigma$ errors are estimated from Fig.2(c)-(e).}

\end{document}